\documentclass[preprint,showpacs,preprintnumbers,amsmath,amssymb]{revtex4}

\usepackage{latexsym}
\usepackage{hyperref}

\begin{document}


\title{Scalar cosmological perturbation in an inflationary brane world
driven by the bulk inflaton} 
\author{Da-Ping Du} 
\email{022019004@fudan.edu.cn}
 \author{Bin Wang} 
 \email{binwang@fudan.ac.cn}
\affiliation{
Department of Physics, Fudan University, Shanghai 200433, P.
R. China}
\author{Elcio Abdalla} 
\email{eabdalla@fma.if.usp.br}
\affiliation{
Instituto de Fisica, Universidade de Sao Paulo,
C.P.66.318, CEP 05315-970, Sao Paulo, Brazil}
\author{Ru-keng Su}
\email{rksu@fudan.ac.cn}
\affiliation{
China Center of Advanced Science and Technology (World Lab), Beijing
100080, and 
Department of Physics, Fudan University, Shanghai 200433, P.
R. China}

\begin{abstract}
We investigate scalar perturbations from inflation in a bulk
inflaton braneworld model. Using the generalized longitudinal
gauge, we derive and solve the full set of scalar perturbation
equations. Our exact results support the recent argument that for
the de Sitter brane the square of the radion mass is not positive,
showing that unlike the flat brane case, the de Sitter brane is
not stable.
\end{abstract}


\maketitle


\section{Introduction}

Based mostly upon string theory, the brane world scenario emerged
in the past few years and offered dramatic changes in our thinking
about the universe (for a review see \cite{1} and references
therein). It is widely believed that we may live on a hypersurface
in higher dimensions with ordinary matter fields being confined on
the brane and graviton propagating through extra dimensions. For
the brane world scenario to become mature and successful, it has
to confront experimental tests. Cosmological perturbations may be
the main avenue to probe the brane world models. The recent
accurate CMB experiment and expected promising future precision
experiments will provide a great deal of information on the
cosmological perturbations and test the brane world model
observationally. With this motivation, theoretically it is of
great interest to clarify the cosmological perturbations in brane
worlds.

There have been a lot of papers studying static geometries with
branes, including flat stabilized branes as well as curved de
Sitter branes. The theory of scalar fluctuations around flat
stabilized branes, involving bulk scalar field fluctuations,
scalar bulk metric fluctuations and brane displacement is well
understood \cite{2}\cite{3}\cite{4}\cite{5}. The system of
dynamical equations can be diagonalized and the extra dimensional
contribution can be separated out. The problem is reduced to
solving a second-order differential equation of the extra
dimensional contribution to the fluctuation satisfying the
boundary condition at the brane. The massive spectrum of the
scalar perturbation appears due to the extra dimensions. Tensor
perturbations of the bulk geometry with stabilized branes have
also been studied (see \cite{6} for a general discussion).

The bulk geometry with curved de Sitter branes can be used to
explain the inflation in the early universe. Inflation can occur
by the dynamics of inflaton either on the brane or in the bulk.
The theory of metric fluctuations around bulk geometry with
inflating branes is more complicated than that for the flat
branes. For the tensor fluctuations, only massless modes can be
generated from inflation \cite{7}. The study of the scalar
perturbations in the braneworld inflation driven by the inflaton
on the brane has been carried out by many authors \cite{8} and it
was found that inflation occurs basically in the same way as that
of the ordinary four-dimensional universe. For the brane inflation
driven by the bulk inflaton, the investigation is rather
complicated. This is partly because the back-reaction to the bulk
geometry should be included in treating the dynamics of the bulk
inflation \cite{9}, and partly because the bulk perturbation
should be taken into account and the full five-dimensional
Einstein equations have to be solved. In many works, perturbations
of the inflation were considered by neglecting metric
perturbations \cite{10}. Recently, progress on this issue has been
achieved. In \cite{11}, the cosmological perturbations in bulk
inflaton model has been investigated by using a covariant
curvature formalism. Further, choosing a specific bulk inflaton
model with an exponential potential and dilatonic coupling to the
brane tension, exact analytic solutions for the scalar
perturbations have been obtained \cite{12}.

In this paper we are going to extend the study set up in \cite{12}
to another bulk inflaton model with a tachyonic bulk potential
having a maximum at $\phi=0$ and without coupling to brane tension
\cite{13}. With the generalized longitudinal gauge \cite{2} for
scalar perturbations, we derive the full set of scalar
perturbation equations in the bulk inflaton model and obtain their
analytic solutions. Using these analytic solutions, we find that
in contrast to the flat brane case, for de Sitter branes the
square of the eigenvalue of the scalar fluctuations is typically
negative. This result was first got by a full nonlinear numerical
treatment \cite{14} and subsequently confirmed by a subsequent
analysis \cite{15}. Without taking any simplification on boundary
conditions and bulk perturbation equations, our exact analytic
result for the specific model can be used as a support to these
arguments.

Our paper is organized as follow. In Section 2 we briefly review
the geometrical background. In section 3 we give the full set of
scalar perturbation equations and the boundary conditions. In
Section 4 we solve the equations analytically. The last section is
devoted to the summary and discussions.  For convenience we also
give some detailed derivations in the appendix.

Throughout the paper we adopt the conventions as follows: Capital
Latin indices mark the bulk, $A,B,C\cdots=0,1,2,3,5$(here 5
represent the extra dimension). The usual four dimensional
space-time is signed by Greek indices running from 0 to 3. And we
use lower case Latin indices ($i,j,k\cdots=1,2,3$) for 3-brane. We
denote the bulk coordinate by $X^A=(t,x^i,r)$ and use the metric
signature $(+,-,-,-,-)$.

\section{Background}

We consider a spacetime with a negative cosmological
constant $\Lambda_5$ as well as a small scalar field $\phi$
inhabited in the bulk while the single brane is empty except for a
positive tension $\sigma_0$ which is minimally coupled with the
scalar field. We begin with an action given by \cite{13}
\begin{equation}\label{eq:action}
\mathcal{S}=\int d^5x
\sqrt{|g|}\Big(-\frac{1}{2k_{5}^2}(\mathcal{R}+2\Lambda_5)
+\frac{1}{2}(\nabla^C\phi)^2-V(\phi)\Big)-\int
d^4x\sqrt{|\gamma|}\sigma_0
\end{equation}
here $k^2_5$ is the five-dimensional gravity constant. To recover
the Randall-Sundrum model, we need to tune the tension
$\sigma_0=\sqrt{-6\Lambda_5/k^4_5}$ in the absence of the bulk scalar
field. The four-dimensional hypersurface locates at
$r=r_0$ and the induced metric reads
\[\gamma_{\mu\nu}=\partial_{\mu}X^A\partial_{\nu}X^Bg_{AB}=\delta^A_\mu\delta^B_\nu
g_{AB}
\]

If we set $k^2_5=1$, the Einstein equation is
\begin{eqnarray}\label{eq:einstein}
{G^A}_B-\Lambda_5{g^A}_B=
\partial^A\phi\partial_B\phi-{g^A}_{B}\Big(\frac{1}{2}(\nabla^C\phi)^2-V(\phi)\Big
)\nonumber\\+\frac{\sqrt{|\gamma|}}{\sqrt{|g|}}\gamma_{\mu\nu}\delta^\mu_B\delta^\nu_C
g^{AC}\sigma_0\delta(r-r_0)
\end{eqnarray}
and the scalar field equation is
\begin{eqnarray}\label{fieldmotion}
\nabla_C\nabla^C\phi+\frac{\partial V(\phi)}{\partial \phi}=0
\end{eqnarray}

In this work, we will concentrate on the potential taking a
tachyonic form $V(\phi)=V_0+\frac{1}{2}M^2\phi^2$, with $V_0>0,
M^2<0$ in the vicinity of $\phi=0$. There exists a maximum of the
potential somewhere at $\phi=\phi_{min}\approx 0$ at which
$V(\phi_{min})=V_0$, where the Randall-Sundrum(RS) type II flat
brane is recovered if $V_0=0$. The non-vanishing $V_0$ would lead to
the deviation from RS model. We assume that $V$ is positive and may
vary very slowly in space and time. The sufficiently slowly
varying of the bulk scalar field will lead to the standard
slow-roll inflation. The effective bulk cosmological constant is
$\Lambda_{5,eff}=\Lambda_5+V_0$ and we need $
V_0<\vert\Lambda_5\vert$ to ensure $\Lambda_{5,eff}<0$. The
curvature radius is $\ell=\sqrt{-\Lambda_{5,eff}/6}$. The bulk
metric is
\begin{eqnarray}
ds^2&=&g_{AB}dX^AdX^B\nonumber\\
 &=&(H\ell)^2\sinh^2y\Big(dt^2-H^{-2}e^{2Ht}\delta_{ij}dx^i
dx^j\Big)-\ell^2dy^2  \quad(y\leq y_0)\label{eq:metric1}
\end{eqnarray}
where we have defined $y=r/\ell$ for simplicity. As a direct
result, the brane position is determined by \cite{13}
\begin{equation}\label{eq:hubble}
H(\ell)=\frac{1}{\ell\sinh y_0}
\end{equation}

At the end of this section we introduce a new coordinate
$\widetilde{X}^A=(u,x^i,v)$. The transformation $X^A\rightarrow
\widetilde{X}^A$ can be realized by
\begin{eqnarray} 
&&u=\frac{\cosh y}{\sinh y}e^{-Ht}\nonumber\\
&&v=\frac{1}{\sinh y}e^{-Ht}\quad ,\label{uvxi} \label{eq:transform} \\ 
&&x^i=x^i\nonumber
\end{eqnarray}
We can rewrite the background metric (\ref{eq:metric1}) into the new form
\begin{equation}\label{metric2}
ds^2=a^2(v)\Big[du^2-\delta_{ij}dx^idx^j-dv^2\Big]
\end{equation}
where we have defined $a^2(v)\equiv (\ell/v)^2$. In the
pseudo-spacetime $\widetilde{X}^A$, it seems just like a RS world
except that the brane is "moving". The simpler geometry in the new
coordinate $\widetilde{X}^A$ will be shown convenient to solve the
perturbations in the bulk since it is possible to diagonalize the
perturbation equations. Such scalar models for a domain wall have
been discussed in the literature in detail also by
\cite{chamreall} and in another context by \cite{abdcua}.

\section{scalar perturbation equations and boundary conditions}

The scalar perturbations in brane world inflation with bulk scalar
is very complicated. Indeed, one has to consider five dimensional
scalar metric fluctuations and brane displacement induced by the
bulk scalar field fluctuation. Taking the generalized longitudinal
gauge for scalar perturbations, we have the perturbed metric in
different coordinate systems $X^A$ and $\widetilde{X}^A$. In the
real spacetime $X^A=(t,x^i,y)$, the perturbed metric can be
written as
\begin{eqnarray}
ds^2=(H\ell)^2\sinh^2y\Big[(1+2\Phi)dt^2-H^{-2}e^{2Ht}(1-2\Psi)\delta_{ij}dx^i
dx^j\Big]\nonumber\\
 -2B\ell dt dy-\ell^2(1-2N)dy^2\quad , \label{eq:pertur1}
\end{eqnarray}
while in the pseudo-spacetime $\widetilde{X}^A=(u,x^i,v)$, we
have the perturbed metric
\begin{eqnarray}
ds^2=a^2(v)\Big[(1+2\varphi)du^2-2Wdudv-(1-2\psi)
\delta_{ij}dx^idx^j-(1-2\Gamma)dv^2\Big]\quad . \label{eq:pertur2}
\end{eqnarray}

By using (\ref{eq:transform}), it is easy to find that the relations
between $(\Phi,\Psi,B,N)$ and $(\varphi,\psi,W,\Gamma)$ are
\begin{eqnarray}\label{eq:relations}
&&\Phi=\coth^2 y\varphi+\frac{1}{\sinh^2 y}\Gamma-\frac{\cosh
y}{\sinh^2 y}W\quad , \nonumber\\
&&N=\frac{1}{\sinh^2 y}\varphi+\coth^2 y\Gamma-\frac{\cosh
y}{\sinh^2 y}W\quad , \label{eq:relationsb}\\
&&B=\frac{1+\cosh^2y}{\sinh y}H\ell W-2\coth y
H\ell(\varphi+\Gamma)\quad , \nonumber\\
&&\Psi=\psi\quad . \nonumber
\end{eqnarray}

After rescaling the five-dimensional cosmological constant, we can
take the perturbed source term $\delta {T^A}_B$ as well as the energy momentum
tensor itself ${T^A}_B$ vanishing at the lowest order
approximation. So the perturbed Einstein equations are simplified to be
\begin{equation}\label{eq:peinstein}
\delta {G^A}_B=0
\end{equation}
Although Einstein equations (\ref{eq:peinstein}) are independent
of coordinate, the boundary conditions(or the differentiation) may
differ in coordinate $X^A$ and $\widetilde{X}^A$. In the coordinate
system $\widetilde{X}^A$, due to the maximally symmetric bulk
spacetime (\ref{metric2}), it is possible to find general solutions for
perturbations in the bulk. As shown in \cite{3}\cite{12}, it can
make perturbation equations diagonal and find general solutions easily.

\subsection{Perturbation Equations}

In this subsection we will work in the coordinate system $\widetilde{X}^A$
in order to derive perturbation equations. In the absence of bulk sources 
with anisotropic stresses, the vanishing of $\delta {G^i}_j$ $(i\neq j)$ 
leads to
\begin{equation}\label{eq:mequation1}
\varphi=\psi+\Gamma\quad .
\end{equation}
Combining the $(i,0)$, $(5,0)$ and $(5,i)$ components of Einstein
equations, we have the equation for $W$,
\begin{equation}\label{eq:w1}
a^2\Box_5W=3\big(\frac{a''}{a}-\frac{a'^2}{a^2}\big)W\quad ,
\end{equation}
where a dot represents the derivative with respect to $u$, and a prime a
derivative with respect to $v$. The five-dimensional d'Alembert operator 
is now
\[
\Box_5=\frac{1}{\sqrt{|g|}}\frac{\partial}{\partial
x^A}\Big(\sqrt{|g|}g^{AB}\frac{\partial}{\partial x
^B}\Big)=\frac{1}{a^2}\Big(\partial^2_u-\partial^2_v-3\frac{a'}{a}
\partial_v-\nabla^2\Big) \quad .
\]

The other equations, namely $(0,0)$, $(i,i)$ and $(5,5)$ can be written as
\begin{eqnarray*}
&&-3a^2\Box_5\psi+3\ddot{\psi}-\nabla^2\psi+\nabla^2\Gamma-3
\frac{a'}{a}\Gamma'-6\frac{a''}{a}\Gamma=0\quad ,\\
&&-a^2\Box_5\psi+a^2\Box_5\Gamma+3\ddot{\psi}-\nabla^2\psi+\nabla^2
\Gamma-3\frac{a'}{a}\Gamma'-6\frac{a''}{a}\Gamma=0\quad ,\\
&&-3\ddot{\psi}+\nabla^2\psi-\nabla^2\Gamma+3\frac{a'}{a}
\Gamma'+6\frac{a'^2}{a^2}\Gamma+6\frac{a'}{a}\phi'\delta\phi=0\quad .
\end{eqnarray*}
They lead to the decoupled equations
\begin{eqnarray}
a^2\Box_5(2\psi+\Gamma)=0\label{eq:psi}
\end{eqnarray}
and
\begin{eqnarray}
a^2\Box_5\Gamma=4\Big(\frac{a''}{a}-\frac{a'^2}{a^2}\Big)\Gamma\label{eq:gama}
\end{eqnarray}
Equations (\ref{eq:w1}), (\ref{eq:psi}) and (\ref{eq:gama}) can be 
expressed in a compact form as
\begin{eqnarray}\label{eq:uniteform}
\Box_5\omega_i=-M_i^2\omega_i\quad ,
\end{eqnarray}
where we introduced the fields
\begin{eqnarray*}
\omega_1=W,&\omega_2=2\psi+\Gamma, & \omega_3=\Gamma
\end{eqnarray*}
where $-M^2_i\ell^2=3,0,4$ for $i=1,2,3$, respectively.

Together, the ``constraint equations", that is  $(i,0), (5,0)$ and $(5,i)$
components of Einstein equations, (\ref{eq:uniteform}) describe the complete
evolution of the scalar fluctuation in coordinate system
$\widetilde{X}^A$.

\subsection{Boundary Conditions}

Perturbing a (bulk) metric in the presence of a brane, we have to
care about obeying the boundary conditions. In fact, any solution
of Einstein Equations in the bulk, in the presence of a brane,
should obey the so-called Darmois-Israel conditions
\cite{israeldarmois}, which in the case of a scalar interaction
was discussed in \cite{abdcua}. Perturbations of such solutions
should be subject to the above boundary conditions. For a static
brane the solution is simple, because the position of the brane is
fixed and generally independent of the brane position. However,
when the brane is dynamical, that is, view from the point of view
of the bulk, we may have distortions of the brane quite difficult
to deal with, since the position of the brane itself depends on
the brane coordinates (see \cite{xx}\cite{abdcas}).

In the present case, if we write boundary conditions in terms of
$\widetilde{X}^A$, that make the bulk equations diagonal, the
boundary conditions will not be diagonal. This is contrary to the
usual static brane where the bulk equations and boundary
conditions can be diagonalized for the same variables \cite{12}. The
complexity of the boundary conditions reflects the fact that our
brane is moving in the coordinate $\widetilde{X}^A$. The boundary
conditions can be diagonalized in the original coordinate
$X^A=(t,x^i,y)$. It was shown that to keep the whole metric to be
effectively $Z_2$ symmetric, the diagonal variables (or $\Phi$,
$\Psi$ and $N$) should be even while the off-diagonal variable (or
$B$) should be odd across the brane \cite{2}. Therefore, from the
perturbed Einstein equations in the coordinate system $X^A$, we
get the boundary conditions
\begin{eqnarray}
&&\Psi'(y_0)=-\Phi'(y_0)=\beta N(y_0)\label{eq:condit11}\\
&&B\big|_{y=y_0}=0 \label{eq:condit12}
\end{eqnarray}
where $\beta=\frac{2\cosh y_0 }{\sinh y_0}-\frac{1}{6}\ell\sigma_0$ for 
a very slowly variation of the bulk scalar field. Notice that the first
derivative of the even functions will give rise to a jump across the
brane, thus its value may have a sign ambiguity. Here we evaluate them in
the positive interval $0\leq y\leq y_0$.

Combining (\ref{eq:hubble}) and
\[
H^2=\frac{1}{6}(\Lambda_{5,eff}+\frac{1}{6}\sigma^2_0)=
-\frac{1}{\ell^2}+\frac{1}{36}\sigma^2_0\quad , 
\]
we get
\[
\beta=\frac{\cosh y_0 }{\sinh y_0}\quad .
\]

\section{Solution of the Perturbed equations}

In order to find solutions which satisfy the boundary conditions at the
brane, it is convenient to do the calculations of the bulk
equations in coordinate $X^A$. This can be simply done by changing
the d'Alembert operator into
\begin{equation*}
\Box_5=\frac{1}{\ell^2\sinh^2y}\Big[\frac{1}{H^2}\partial^2_t+
\frac{3}{H}\partial_t-\sinh^2y\partial^2_y-4\sinh
y\cosh y\partial_y-e^{-2Ht}\nabla^2\Big]\quad .
\end{equation*}

Separating (\ref{eq:uniteform}) by
\[
\omega=\int\textrm{d}\lambda\,
u_\lambda(y)\omega_\lambda(t,\textbf{x})\quad , 
\] 
we have the equations
\begin{equation}\label{former}
  u''_\lambda+4\coth
y\,u'_\lambda+(-M^2_i\ell^2+\frac{\lambda^2}{\sinh^2y})u_\lambda=0\
\end{equation}
and
\begin{equation}\label{latter}
H^2(\Box_4+\lambda^2)\omega_\lambda=\ddot{\omega}_\lambda
+3H\dot{\omega}_\lambda+(-e^{-2Ht}\nabla^2+\lambda^2)H^2
\omega_\lambda=0\quad .
\end{equation}

Above, $\lambda^2$ is the separation constant corresponds to the
Kaluza-Klein (KK) mass which appears in the KK compactification.
The general solution for equation (\ref{latter}) is
\[
\omega_\lambda=\frac{1}{(2\pi)^{3/2}}\int \textrm{d}^3p\;\;
(-\eta)^{3/2}\mathcal{H}_\nu(-p\eta)e^{ip\textrm{x}} \quad ,
\]
where $\nu^2=(9/4)-\lambda^2$, and we have expressed it with the
conformal time $\eta=-e^{-Ht}$. $\mathcal{H}_\nu$ is the arbitrary
combination of the Hankel functions $\mathcal{H}^{(1)}$ and
$\mathcal{H}^{(2)}$.

The general solution for equation (\ref{former}) is
\[
u_\lambda(y)=A_1\frac{P^\nu_{\mu-1/2}(\cosh y)}{\sinh^{3/2}y}+
A_2\frac{Q^\nu_{\mu-1/2}(\cosh y)}{\sinh^{3/2}y}\quad ,
\]
where $\mu_i=1,2,0$ for $i=1,2,3$, respectively.

In the coordinate system $\widetilde{X}^A$ the solutions $\omega_i$ should
satisfy ``constraint equations", that is the $(i,0)$, $(5,i)$ and $(5,0)$
components of Einstein equations. Expressing the first two constraint
equations, namely the $(i,0)$ and $(5,i)$ components of Einstein equations
in terms of coordinates  $X^A(t,x^i,y)$, we have 
\begin{eqnarray}
&&\cosh y\sinh y\omega'_1-2\sinh
y\omega'_2+3\sinh^2y\omega_1=\frac{1}{H}(\dot{\omega}_1-2\cosh
y\dot{\omega}_2)\quad , \label{eq:constrain1}\\
&&\sinh y\omega'_1+\cosh y\sinh
y(-3\omega'_3+\omega'_2)-6\sinh^2y\omega_3=\frac{1}{H}(\cosh
y\dot{\omega}_1-3\dot{\omega}_3+\dot{\omega}_2)\quad .\label{eq:constrain2}
\end{eqnarray}
where a dot (or a prime) donates the derivative with $t$ (or $y$).

If we get a set of solutions of $\omega_i$, the results
for $(\Phi,\Psi,B,N)$ can be obtained immediately through  relations
(\ref{eq:relations}). In the rest of this section, we
focus on the possible solutions for $\omega_i$.

Constraint (\ref{eq:constrain1}) involves only $\omega_1$ and
$\omega_2$. With the property of the Hankel function, one notices
that the derivative with respect to the conformal time $\eta$ will give
rise to a term with a factor $p\eta$ which cannot be cancelled by
terms on the left side of (\ref{eq:constrain1}). Therefore, the solutions
for $\omega_1$ and $\omega_2$ should contain the
$\mathcal{H}_{\nu-2}$ term as well as the $\mathcal{H}_{\nu}$
term. We assume that
\begin{eqnarray*}
\omega_1(\nu)=C_1\frac{R^\nu_{1/2}(\cosh
y)}{\sinh^{3/2}y}(-\eta)^{3/2}\mathcal{H}_\nu(-p\eta)+
D_1\frac{R^{\nu-2}_{1/2}(\cosh
y)}{\sinh^{3/2}y}(-\eta)^{3/2}\mathcal{H}_{\nu-2}(-p\eta)\quad , \\
\omega_2(\nu)=C_2\frac{R^\nu_{3/2}(\cosh
y)}{\sinh^{3/2}y}(-\eta)^{3/2}\mathcal{H}_\nu(-p\eta)
+D_2\frac{R^{\nu-2}_{3/2}(\cosh
y)}{\sinh^{3/2}y}(-\eta)^{3/2}\mathcal{H}_{\nu-2}(-p\eta)\quad , 
\end{eqnarray*}
where $R^\nu_\mu$ is an arbitrary combination of associative
Legendre function $P^\nu_\mu$ and $Q^\nu_\mu$.

With the constraint (\ref{eq:constrain1}), we found that the
coefficients should satisfy (see Appendix(\ref{sec:match2constrain1}) 
for a detailed derivation)
\begin{eqnarray}
&&C_1(\nu)=h_\nu\quad , \nonumber\\
&&D_1(\nu)=(\frac{5}{2}-\nu)(\nu-\frac{1}{2})h_\nu\quad ,
\label{eq:coefficient1} \\ 
&&C_2(\nu)=\frac{1}{2}h_\nu\quad ,\nonumber\\
&&D_2(\nu)=\frac{1}{2}(\frac{5}{2}-\nu)(\frac{7}{2}-\nu)h_\nu\quad , \nonumber
\end{eqnarray}
where $h_\nu$ is an arbitrary value.

However, not all the solutions that satisfy the constraint
(\ref{eq:constrain1}) can meet the requirement (\ref{eq:constrain2}). 
One is always allowed to expand the target solutions in the series of 
solutions that satisfy (\ref{eq:constrain1}), that is
\begin{eqnarray}
&&\omega_1=\sum^{n=+\infty}_{n=-\infty}a_{\nu_0+2n}
\frac{R^{\nu_0+2n}_{1/2}}{\sinh^{3/2}y}(-\eta)^{3/2}
\mathcal{H}_{\nu_0+2n}(-p\eta)=\sum^{n=+\infty}_{n=-\infty}
\omega_1(\nu_0+2n)\quad , \nonumber\\
&&\omega_2=\sum^{n=+\infty}_{n=-\infty}b_{\nu_0+2n}
\frac{R^{\nu_0+2n}_{3/2}}{\sinh^{3/2}y}(-\eta)^{3/2}
\mathcal{H}_{\nu_0+2n}(-p\eta)=\sum^{n=+\infty}_{n=-\infty}
\omega_2(\nu_0+2n)\quad ,\label{eq:assumption}  \\
&&\omega_3=\sum^{n=+\infty}_{n=-\infty}c_{\nu_0+2n}
\frac{R^{\nu_0+2n}_{-1/2}}{\sinh^{3/2}y}(-\eta)^{3/2}
\mathcal{H}_{\nu_0+2n}(-p\eta)\nonumber
\end{eqnarray}
where $\nu_0$ is arbitrary and $\omega_1(\nu)$ and $\omega_2(\nu)$
are defined as above with coefficients determined as
in (\ref{eq:coefficient1}). Note that here $R^\nu_\mu$ is not the
same as that in $\omega_i(\nu)$ since the coefficients of
$P^\nu_\mu$ and $Q^\nu_\mu$ may be different.

By imposing the constraint (\ref{eq:constrain2}), we found
that (see Appendix (\ref{sec:match2constrain2}))
\begin{eqnarray}
&&a_{\nu}=-\frac{4\nu h_{\nu}}{(\frac{5}{2}+\nu)(\frac{3}{2}-\nu)}\nonumber\\
&&b_{\nu}=\frac{2\nu h_{\nu}}{(\frac{5}{2}+\nu)}\quad ,
\label{eq:coefficient2} \\ 
&&c_{\nu}=-\frac{2}{3}h_\nu\frac{\nu(\frac{1}{2}+\nu)}{(\frac{5}{2}+
\nu)(\frac{3}{2}-\nu)}\quad , \nonumber
\end{eqnarray}
where $h_\nu$ is defined by the recursion relation
\begin{equation}\label{eq:h}
\frac{h_{\nu}}{h_{\nu+2}}=-(\frac{1}{2}-\nu)(\frac{3}{2}-\nu)
\frac{\frac{5}{2}+\nu}{\frac{5}{2}-\nu}\quad .
\end{equation}

Although we have not yet imposed any constraint on the effective mass
$\lambda^2$ (related to the index $\nu$), one can easily find that in order
to make the solutions physically appropriate we should remove the
divergent part of the summation (\ref{eq:assumption}). A cutoff will
occur when the recursion meets zero. There are only two possible solutions.

In one case, we find from the relation (\ref{eq:h}) that $h_{1/2}=0$
when $\nu_0=1/2$. This leads to the fact that all $h_{\nu}$ with 
$\nu=\nu_0-2\vert n\vert$ will vanish. By inverting the relation 
(\ref{eq:h}), one finds that the terms above $h_{5/2}$ with $\nu=5/2+2
\vert n\vert$ also vanish. The only nonzero term in the series expansion 
is $\nu=5/2$.

In another case, the series is safely cutoff at $\nu_0=3/2$.
$h_{3/2}=0$ (so are the lower terms) if $h_{7/2}$ is finite. The
terms above $h_{7/2}$ are all physical. It should be noted
that $a_{3/2}$ and $c_{3/2}$ will not diverge for $h_{3/2}=0$. The
solutions contain infinite terms and have the lowest index at $\nu_0=3/2$.

We can express the solutions explicitly. In the first case
\begin{eqnarray}
&&\omega_1=2h\;\frac{R^{5/2}_{1/2}}{\sinh^{3/2}y}(-\eta)^{3/2}
\mathcal{H}_{5/2}(-p\eta)\quad , \nonumber\\
&&\omega_2=h\frac{R^{5/2}_{3/2}}{\sinh^{3/2}y}(-\eta)^{3/2}
\mathcal{H}_{5/2}(-p\eta)\quad , \label{eq:solution1} \\
&&\omega_3=h\frac{R^{5/2}_{-1/2}}{\sinh^{3/2}y}(-\eta)^{3/2}
\mathcal{H}_{5/2}(-p\eta)\quad , \nonumber
\end{eqnarray}
In the second case
\begin{eqnarray}
&&\omega_1=\sum^{+\infty}_{k=0}\,-h\frac{\frac{16}{3}(
\frac{3}{2}+2k)}{(2k-1)(2k)!![2(k+2)]!!}\;\;
\frac{R^{3/2+2k}_{1/2}}{\sinh^{3/2}y}(-\eta)^{3/2}
\mathcal{H}_{3/2+2k}(-p\eta)\quad , \nonumber\\
&&\omega_2=\sum^{+\infty}_{k=0}\,-h\frac{\frac{8}{3}(
\frac{3}{2}+2k)}{(2k-1)[2(k-1)]!![2(k+2)]!!}\;\;
\frac{R^{3/2+2k}_{3/2}}{\sinh^{3/2}y}(-\eta)^{3/2}
\mathcal{H}_{3/2+2k}(-p\eta)\quad ,  \label{eq:solution2}\\
&&\omega_3=\sum^{+\infty}_{k=0}\,-h\frac{\frac{8}{9}(
\frac{3}{2}+2k)(2k+2)}{(2k-1)(2k)!![2(k+2)]!!}\;\;
\frac{R^{3/2+2k}_{-1/2}}{\sinh^{3/2}y}(-\eta)^{3/2}
\mathcal{H}_{3/2+2k}(-p\eta)\quad , \nonumber
\end{eqnarray}
where $h$ is arbitrary and we have defined
\[
(2n)!!=\prod^{n}_{k=1}(2k)\quad , 
\]
with the convention that $0!!=1$ and $(-2)!!=\infty$.

Notice that $\omega_1$ is the solution of the combination of
constraint equations, so the other constraint given by $(5,0)$
equation will be satisfied  automatically. These are all possible
solutions for $\omega_i$ satisfying all the constraint equations.
Hence they are the exact solutions for the Einstein equations.

The remaining task is to test whether the solutions obtained can satisfy
the boundary condition to make the metric effectively $Z_2$ symmetric. 
We will restrict ourselves in the case $H\ell\ll 1$, or equivalently
$\sinh y_0\sim\cosh y_0\gg 1$.

In (\ref{eq:assumption})
\[
R^{\nu}_\mu(y)=P^{\nu}_\mu(y)-\alpha_\nu Q^{\nu}_\mu(y)
\]
where $\mu=1/2,3/2,-1/2$. The coefficient $\alpha_\nu$ will be
determined from the condition (\ref{eq:condit12}).

It is worthwhile mentioning that in the background we are
interested in, the solutions indeed satisfy condition
(\ref{eq:condit11}), thus the perturbed background is
effectively $Z_2$-symmetric (see Appendix (\ref{sec:fixcondition})).

\section{summary and discussions}

In this paper, we derived the exact analytic solutions for scalar
perturbations in a brane world model with curved de Sitter brane.
Due to the existence of the bulk scalar field, the scalar
perturbations are quite complicated. We found a coordinate system
which can make the perturbation equations in the bulk diagonal. We
have obtained the analytic solutions satisfying boundary
conditions and other constraint equations.

In order to make the result physically legal, we have to cut off the
divergent part of the solutions by restricting the index $\nu$. This leads
to two results: for the first, $\nu=5/2$, the series contain only one term
as shown in (\ref{eq:solution1}); for the second case, the solution is a
combination of infinite waves as shown in (\ref{eq:solution2}) and the
index has a lowest value at $\nu=3/2$. In both cases the index of the
waves, $\nu$, is real. Considering $\nu^2=9/4-\lambda^2$, the requirement
$\nu\geq 3/2$ leads to the KK mass $\lambda^2\leq 0$, implying the the
solution involving the tachyonic mode is instable. This result agrees to
the numerical argument, that different from the flat brane case, for the
de Sitter branes the square of the radion mass is not positive, which
leads to a strong tachyonic instability \cite{14}\cite{15}. Our exact
analytic result can be used as a support to these argument. 

The tachyonic instability for inflating branes means that the
braneworlds with inflation driven by the bulk inflaton are hard to
stabilize. It is of interest to study what kind of mechanism can
be introduced to stabilize the brane world. One expectation is
that the stabilization can be achieved by introducing another
scalar field on the brane in addition to the one in the bulk \cite{15}.
Whether this attempt can work needs further investigation.

\appendix
\section{scalar perturbation of the metric}\label{sec:pmetric}

In the pseudo spacetime $\widetilde{X}^A=(u,x^i,v)$,
\begin{eqnarray*}
&&a^2\delta {G^0}_0=\Big[2\nabla^2+3\frac{\partial^2}{\partial
v^2}+9\frac{a'}{a}\frac{\partial}{\partial
v}\Big]\psi+\Big[\nabla^2-3\frac{a'}{a}\frac{\partial}{\partial
v}-6\frac{a''}{a}\Big]\Gamma\\
&&a^2\delta {G^i}_j=\Big[\nabla^2-\frac{\partial^2}{\partial
v^2}-3\frac{a'}{a}\frac{\partial}{\partial
v}\Big]\varphi+\Big[-\nabla^2+2\frac{\partial^2}{\partial
v^2}+6\frac{a'}{a}\frac{\partial}{\partial
v}-2\frac{\partial^2}{\partial
u^2}\Big]\psi\nonumber\\&&\qquad+\Big[-\nabla^2-3\frac{a'}{a}
\frac{\partial}{\partial
v}-6\frac{a''}{a}-\frac{\partial^2}{\partial
u^2}\Big]\Gamma+\Big[-\frac{\partial^2}{\partial u\partial
v}-3\frac{a'}{a}\frac{\partial}{\partial u}\Big]W \nonumber\\&&
\qquad\qquad(i=j)\\
&&a^2\delta {G^i}_j=\partial_i\partial_j(\varphi-\psi-\Gamma)
\qquad\qquad (i\neq j)\\
&&a^2\delta
{G^5}_5=\Big[-\nabla^2-3\frac{a'}{a}\frac{\partial}{\partial
v}\Big]\varphi+\Big[2\nabla^2+9\frac{a'}{a}\frac{\partial}{\partial
v}-3\frac{\partial^2}{\partial
u^2}\Big]\psi\nonumber\\&&\qquad-3\frac{a'}{a}\dot{W}-12
\frac{a'^2}{a^2}\Gamma\\
&&a^2\delta {G^i}_0=\partial_i\Big(-\frac{W'}{2}-\frac{3}{2}
\frac{a'}{a}W-2\dot{\psi}-\dot{\Gamma}\Big)\\
&&a^2\delta
{G^5}_0=\frac{1}{2}\nabla^2W-3\dot{\psi'}+3\frac{a'}{a}\dot{\Gamma}\\
&&a^2\delta
{G^5}_i=\partial_i\Big(\frac{\dot{W}}{2}+\varphi'-2\psi'+3
\frac{a'}{a}\Gamma\Big)
\end{eqnarray*}
where a dot represents one derivative over $u$, while a prime over
$v$.

\section{Matching the constraints}
\subsection{Matching (\ref{eq:constrain1})}\label{sec:match2constrain1}

The following recurrence relations of Legendre
functions \cite{handbook} are useful for the derivations
\begin{eqnarray}
&&R^{\nu+1}_\mu(z)=(z^2-1)^{-1/2}\Big[(\mu-\nu)z\,R^\nu_\mu(z)
-(\mu+\nu)R^\nu_{\mu-1}(z)\Big]\label{fm:L1}\\
&&(\mu-\nu+1)R^{\nu}_{\mu+1}(z)=(2\mu+1)zR^{\nu}_\mu(z)-
(\mu+\nu)R^{\nu}_{\mu-1}(z)\label{fm:L2}\\
&&(\mu-\nu+1)(z^2-1)^{1/2}R^{\nu-1}_\mu=zR^\nu_\mu-R^\nu_{\mu-1}\label{fm:L3}
\end{eqnarray}
Note that the recurrences are both valid for $P^\nu_\mu$ and
$Q^\nu_\mu$, so we simply use the notation $R^\nu_\mu$ to denote
the former or the latter or the linear combination of them.

With the properties of the Hankel functions we have 
\[
\frac{\textrm{d}}{\textrm{d}t}((-\eta)^{3/2}\mathcal{H}_\nu(-p\eta))
=H\eta\Big[\,p(-\eta)^{3/2}\mathcal{H}_{\nu-1}+(
\frac{3}{2}-\nu)(-\eta)^{1/2}\mathcal{H}_{\nu}\Big]\quad ,
\]
\[
\frac{\textrm{d}}{\textrm{d}t}((-\eta)^{3/2}\mathcal{H}_{\nu-2}(-p\eta))
=H\eta\Big[\,-p(-\eta)^{3/2}\mathcal{H}_{\nu-1}+(\nu-\frac{1}{2})
(-\eta)^{1/2}\mathcal{H}_{\nu-2}\Big] \quad .
\]

Considering the right hand side of the constraint (\ref{eq:constrain1}),
the $\mathcal{H}_{\nu-1}$ terms in $\dot{\omega}_1$ and
$\dot{\omega}_2$ should be eliminated, that is (here and
after we will simplify $\underline{\sinh y}$($\underline{\cosh
y}$) as S(C) for short)
\[
\Big[C_1R^\nu_{1/2}(C)-2CC_2R^{\nu}_{3/2}(C)\Big]-
\Big[D_1R^{\nu-2}_{1/2}(C)-2CD_2R^{\nu-2}_{3/2}(C)\Big]=0\quad .
\]
Using (\ref{fm:L3}) for the first term and (\ref{fm:L1}) for
the second term, one can find the desired relations for the
coefficients (\ref{eq:coefficient1}).

We can show that these relations indeed satisfy constraint equation
(\ref{eq:constrain1}). After this setting, what is left at the right
hand side is
\[
\dot{\omega}_1-2C\dot{\omega}_2=h(\frac{5}{2}-\nu)
\frac{SR^{\nu-1}_{3/2}}{S^{3/2}}(-\eta)^{3/2}\Big[(
\frac{3}{2}-\nu)\mathcal{H}_{\nu}+(\nu-\frac{1}{2})\mathcal{H}_{\nu-2}\Big]
\]

With the help of the derivative recurrence of Legendre
functions \cite{handbook}, we can do the derivatives
\begin{eqnarray}
 &&\frac{\textrm{d}}{\textrm{d}y}\Big(\frac{R^\nu_{3/2}(C)}
{S^{3/2}}\Big)=-(\nu+\frac{3}{2})\frac{R^\nu_{1/2}(C)}{S^{5/2}} \quad ,
\label{eq:derive}\\
&&\frac{\textrm{d}}{\textrm{d}y}\Big(\frac{R^\nu_{1/2}(C)}{S^{3/2}}
\Big)=-\frac{3CR^\nu_{1/2}(C)}{S^{5/2}}+(\frac{3}{2}-\nu)
\frac{R^\nu_{3/2}(C)}{S^{5/2}}\quad .
\end{eqnarray}

Therefore, the left hand side of (\ref{eq:constrain1})  becomes
\begin{eqnarray*}
&&CS\omega'_1-2S\omega'_2+3S^2\omega_1\\
&&=h\frac{(-\eta)^{3/2}}{S^{3/2}}\Big\{(\frac{3}{2}-\nu)
\mathcal{H}^\nu\Big[CR^\nu_{3/2}-R^\nu_{1/2}\Big]\qquad\qquad\\
&&\qquad\;+(\nu-\frac{1}{2})\mathcal{H}_{\nu-2}
\Big[(\frac{7}{2}-\nu)(\frac{5}{2}-\nu)CR^{\nu-2}_{3/2}-
(\frac{5}{2}-\nu)(\nu-\frac{1}{2})R^{\nu-2}_{1/2}\Big]\Big\}\quad .
\end{eqnarray*}
Using (\ref{fm:L3}) and (\ref{fm:L1}), we know that this is exactly
the same as that in the right hand side.

\subsection{Matching (\ref{eq:constrain2})}\label{sec:match2constrain2}

Up to now, we have gotten the solutions satisfying
(\ref{eq:constrain1}). Thus, we can write out
\begin{eqnarray*}
&&\omega_1(\nu)=2h_\nu\frac{R^\nu_{1/2}}{S^{3/2}}(-\eta)^{3/2}
\mathcal{H}_\nu(-p\eta)+2h_\nu(\frac{5}{2}-\nu)(\nu-
\frac{1}{2})\frac{R^{\nu-2}_{1/2}}{S^{3/2}}(-\eta)^{3/2}
\mathcal{H}_{\nu-2}(-p\eta)\quad ,\\
&&\omega_2(\nu)=h_\nu\frac{R^\nu_{3/2}}{S^{3/2}}(-\eta)^{3/2}
\mathcal{H}_\nu(-p\eta)+h_\nu(\frac{5}{2}-\nu)(\frac{7}{2}-
\nu)\frac{R^{\nu-2}_{3/2}}{S^{3/2}}(-\eta)^{3/2}
\mathcal{H}_{\nu-2}(-p\eta)\quad ,
\end{eqnarray*}
and
\begin{eqnarray*}
&&\omega_1(\nu+2)=2h_{\nu+2}\frac{R^{\nu+2}_{1/2}}{S^{3/2}}
(-\eta)^{3/2}\mathcal{H}_{\nu+2}(-p\eta)+2h_{\nu+2}(
\frac{1}{2}-\nu)(\nu+\frac{3}{2})\frac{R^{\nu}_{1/2}}
{S^{3/2}}(-\eta)^{3/2}\mathcal{H}_{\nu}(-p\eta)\quad ,\\
&&\omega_2(\nu+2)=h_{\nu+2}\frac{R^{\nu+2}_{3/2}}{S^{3/2}}
(-\eta)^{3/2}\mathcal{H}_{\nu+2}(-p\eta)+h_{\nu+2}(
\frac{1}{2}-\nu)(\frac{3}{2}-\nu)\frac{R^{\nu}_{3/2}}
{S^{3/2}}(-\eta)^{3/2}\mathcal{H}_{\nu}(-p\eta)\quad ,
\end{eqnarray*}
where the ratio of $h_\nu$ and $h_{\nu+2}$ are left to be
determined. According to (\ref{eq:assumption}), it is obvious
that
\begin{eqnarray*}
&&a_\nu=2h_{\nu}+2h_{\nu+2}(\frac{1}{2}-\nu)(\nu+\frac{3}{2})\quad ,\\
&&b_\nu=h_{\nu}+h_{\nu+2}(\frac{1}{2}-\nu)(\frac{3}{2}-\nu)\quad .
\end{eqnarray*}

The right hand side of constraint equation (\ref{eq:constrain2}) becomes
\begin{eqnarray*}
&&- \sum_{\nu}S^{-3/2}\Big\{(\frac{3}{2}-\nu)\Big[a_\nu
CR^{\nu}_{1/2}+b_\nu R^{\nu}_{3/2}-3c_\nu
R^{\nu}_{-1/2}\Big](-\eta)^{3/2}\mathcal{H}_{\nu}\\
&&\qquad\qquad\qquad\qquad+\Big[a_\nu CR^\nu_{1/2}+b_\nu
R^\nu_{3/2}-3c_\nu
R^\nu_{-1/2}\Big]p(-\eta)^{5/2}\mathcal{H}_{\nu-1}\Big\}\quad .
\end{eqnarray*}
The second term should be zero. With (\ref{fm:L2}), we
can determine $a_{\nu}$, $b_{\nu}$ and $c_{\nu}$ as in
(\ref{eq:coefficient2}) and the recurrence (\ref{eq:h}) of $h_\nu$.

Following the same procedure as discussed in
Appendix (\ref{sec:match2constrain1}), the constraint
(\ref{eq:constrain2}) is indeed satisfied.

\section{Fixing the boundaries}\label{sec:fixcondition}

For convenience, we write (\ref{eq:relations})
in term of $\omega_i$
\begin{eqnarray*}
&&\Phi=\frac{1}{2S^2}[C^2\omega_2+(2+C^2)\omega_3-C\omega_1]\quad ,\\
&& N=\frac{1}{2S^2}[\omega_2+(1+2C^2)\omega_3-C\omega_1]\quad ,\\
&&B=\frac{H\ell}{S}[(1+C^2)\omega_1-C\omega_2-3C\omega_3]\quad .\\
&&\Psi=\frac{1}{2}[\omega_2-\omega_3]
\end{eqnarray*}

From Appendix (\ref{sec:match2constrain2}), we know the following
identity will always be true
\[
C\omega_1+\omega_2=3\omega_3\quad .
\]

With (\ref{eq:condit12}) we can further find that on the brane
$\omega_1(y_0)=2C\omega_2(y_0)$. Thus, condition (\ref{eq:condit11})
is equal to
\begin{eqnarray*}
&&\frac{1}{S^2}\Big[S\omega_1+2C^2\omega'_2\Big]_{y=y_0}=0 \\
&&\frac{1}{S^3}\Big[\frac{1}{2}(C^2+1)\omega_1-\frac{1}{3}S^2
\omega_2-\frac{2}{3}S^3\omega'_2\Big]_{y=y_0} = 0\quad .
\end{eqnarray*}

The background can only be stabilized if the metric
perturbations keep in small scale far less than the metric, say
$|\omega_i|\ll 1\ll \sinh y_0$. With (\ref{eq:derive}), we have
\[
\omega'_2(y_0)=-\frac{1}{S}\Big[\sum_{\nu}\;(\nu+\frac{3}{2})b_\nu
\frac{R^\nu_{1/2}}{S^{3/2}}(-\eta)^{3/2}H_\nu(-p\eta)\Big]_{y=y_0}\sim
0\quad ,
\]
(where for (\ref{eq:solution1}) $\nu$ is only $5/2$, while
in (\ref{eq:solution2}) $\nu$ runs from $3/2$ to infinite). Thus
the boundary conditions are satisfied. In other words, the
perturbed background is also effectively $Z_2$-symmetric.

ACKNOWLEDGEMENT: This work was partially supported by NNSF of China,
Ministry of Education of China and Ministry of Science and Technology of
China under Grant NKBRSFG19990754. The work of E. Abdalla was supported by
FAPESP and CNPQ, Brazil.


\end{document}